\documentclass[prb,twocolumn,showpacs,groupedaddress,amsmath,amssymb]{revtex4}
\usepackage{graphicx}
\usepackage{bm}

\begin{document}

\title{Synchronization of high-frequency oscillations of phase-slip centers in
a tin whisker under microwave radiation}
\author{V.~I.~Kuznetsov}
\email[Electronic address:]{kvi@ipmt-hpm.ac.ru}
\author{V.~A.~Tulin}
\affiliation{Institute of Microelectronics Technology and High
Purity Materials, Russian Academy of Sciences, 142432
Chernogolovka, Moskow Region, Russia}

\date{\today}

\begin{abstract}
Current-voltage characteristics of a system with a variable number
of phase-slip centers resulting from phase separation in a tin
whisker under external microwave field with a frequency $\Omega /2
\pi  \simeq 35-45$ GHz have been studied experimentally. Emergence
and disappearance of steps with zero slope in a whisker's
current-voltage characteristic at $U_{m/n}=(m/n)U_{\Omega}$, where
$m$ and $n$ are integers and $U_{\Omega}$ is determined by
Josephson's formula $\hbar \Omega=2eU_{\Omega}$, have been
investigated. Microwave field generated by phase-slip centers is
nonharmonic, and the system of phase-slip centers permits
synchronization of internal oscillations at a microwave frequency
by an external field with a frequency which is the $n$-th harmonic
of internal oscillations. The estimated microwave power generated
by a whisker is $10^{-8}$ W. Stimulation of superconductivity in a
current-carrying whisker has been detected.
\end{abstract}

\pacs{74.40.+k, 74.50.+r, 74.25.Nf, 74.78.-w }
\maketitle

\section{\label{sec:level1}INTRODUCTION}
Microwave generation in a Josephson junction (a weak-coupling
element in a superconducting circuit) under a dc voltage has
attracted researchers' attention since the time when the ac
Josephson effect was discovered. The simple structure of the
experimental device and easy control of the generated frequency
are the most attractive features of the effect. The frequency
generated by the junction is determined by the formula
\begin{equation}
\omega=2Ue/ \hbar \;, \nonumber
\end{equation}
where $U$ is the voltage drop across the junction, $e$ is the
electron charge, and $\hbar$ is the Planck constant. The
disadvantages of these devices are their low output and
difficulties in matching the superconducting circuits containing
Josephson junctions to the microwave circuits. Attempts have been
made to overcome these difficulties using circuits of short
junctions \cite{c1,c2,c3,c4}.

Josephson junctions have a typical linear size in the direction
perpendicular to the supercurrent density vector, namely the
Josephson penetration depth $\lambda_{j}$. If the junction
dimension in the direction perpendicular to the supercurrent
satisfies the condition $d<\lambda_{j}$ (a short junction), the
phase variation is uniform over the junction volume, and one has a
single source of microwave radiation. In the case of a network of
short synchronized junctions, it seems possible to derive a high
microwave output close to the sum of powers generated by each
element.

A long uniform superconducting channel with phase-slip centers can
be classified with such systems. Phase-slip centers occur in
resistive states of a long narrow channel carrying a constant
current at a temperature close to the superconducting transition
($I>I_{c}$, $T<T_{c}$) \cite{c5,c6}. Real structures in which
phase-slip centers have been detected are thin films with width
$w$ and single-crystal wires (whiskers) with diameter $d$ smaller
than the superconductor coherence length $\xi$. From the viewpoint
of experimenters dealing with superconducting channels, whiskers
(thin crystalline wires) are preferable because their uniformity
over the length is higher. But thin films have some advantages
when applications are concerned, since their dimensions are
directly controlled during their manufacture. On the other hand,
microscopic inhomogeneities due to fabrication technologies can
lead to considerable degradation of parameters of phase-slip
centers, and a thin film may behave like a system of weak
superconducting bounds localized along the narrow film.

An isolated phase-slip center is an nonstationary, inhomogeneous
entity "localized in the space" and containing an internal region
with a size of about $\xi$ where the superconducting order
parameter oscillates at the Josephson frequency
\begin{equation}
\omega=2U_{\Omega} e/ \hbar \;. \nonumber
\end{equation}
At temperatures near the transition point, the voltage averaged
over the oscillation period, $U_{\omega}$, in the phase-slip
center is due to penetration of a non-uniform longitudinal
electric field into the outer region of the center through a
distance of about $l_{E}$ (the electric field penetration range),
and the electric resistance of each phase-slip center is
\begin{equation}
R_{0}=2\rho_{N}l_{E}/S \;, \nonumber
\end{equation}
where $\rho_{N}$ is the material resistivity in the normal state
and $S$ is the channel cross section \cite{c5,c6,c7,c8}. At the
moment when the absolute value of the order parameter vanishes,
the phase difference over the center jumps by $2 \pi$.
Current-voltage characteristics (CVC) of such superconducting
channels contain a set of sloped linear sections corresponding to
resistances
\begin{equation}
R_{n}=nR_{0} \;, \nonumber
\end{equation}
(where $n$ is an integer) connected by sections of curves with
current jumps. Extrapolations of these linear sections cross the
current axis at approximately the same point $I_{0}$ (an excess
current) \cite{c5,c6,c7,c9}.

Although the number of publications dedicated to phase-slip
centers is fairly large \cite{c5,c6}, the dynamics of systems with
phase-slip centers has been studied insufficiently
\cite{c7,c8,c10,c11,c12}. The reversed ac Josephson effect under
external electromagnetic radiation was detected in thin tin films
at a frequency of 10 GHz \cite{c7} and in single-crystal wires
(whiskers) at frequencies of up to 900 MHz \cite{c13,c14,c15}. In
both these cases, a CVC contains, in addition to sloping steps, a
fundamental step with a zero slope at voltage $U_{\Omega}$ in the
region of parameters corresponding to one phase-slip center and
associated with high-frequency oscillations of the order parameter
in the center, and "weak" steps at
\[U_{m/n}=(m/n)U_{\Omega} \;,\]
where $U_{\Omega}$ is the voltage corresponding to the external
field frequency and $m$ and $n$ are integers. Ivlev and Kopnin
\cite{c16} analyzed the ac Josephson effect in terms of the
microscopic theory. The pattern of various zero-slope steps at
different direct currents and microwave frequencies in systems
with variable numbers of phase-slip centers has not been
investigated in full. A current-carrying whisker under an
electromagnetic field with a frequency higher than 900 MHz has
never been studied.

\section{SAMPLES AND EXPERIMENTAL DETAILS}
In the reported work, we have studied the effect of microwave
fields with frequencies ranging between 35 and 45 GHz on CVCs of
tin whiskers in the regime when several phase-slip centers exist
in a sample at voltages of order of $U_{\Omega}$. In previous
experiments \cite{c7,c13,c14,c15} the parameters $T$ and $\Omega/2
\pi$ were selected so that the mean voltage $U_{\omega}$ across
one center, which determined the frequency of proper
high-frequency oscillations,
\begin{equation}
\omega=2Ue/ \hbar \;, \nonumber
\end{equation}
could be tuned to $U_{\Omega}$, i.e., the frequency $\omega$ of
internal oscillations should be equal to that of applied microwave
field. We have used higher microwave frequencies $\Omega /2 \pi$
and temperatures at a greater distance from $T_{c}$ than Tidecks
et al. \cite{c13,c14,c15} so that to satisfy the condition
\begin{equation}
U_{\omega}=U_{\Omega}/n \;, \nonumber
\end{equation}
i.e., $n \omega=\Omega$ ($n>1$) when a sample contained several
phase-slip centers at voltages about $U_{\Omega}$. It follows from
the microscopic theory \cite{c16} that this is the condition under
which induced steps on a CVC are generated at voltages
$U_{\Omega}/n$. The presence of such steps means that the
radiation generated by the system of phase-slip centers is
nonharmonic. Given the higher uniformity of whiskers over their
lengths and smaller number of structural defects than in films,
they are preferable for such experiments. Moreover, zero-slope
steps on a CVC of an irradiated whisker \cite{c13,c14,c15} are
considerably wider than in narrow films \cite{c7}. In many
experiments (see for example Ref. \cite{c17}), low-frequency
oscillations instead of high-frequency oscillations were detected
in narrow films. Whiskers grown from thin tin films deposited on
silicon substrates had diameters $d=0.2-0.8$ $\mu$m, lengths of
about 1 mm, resistance ratio $R_{300} /R_{4.2}<100$, and $T_{c}
\approx 3.1$ K. A whisker was set across a 300-$\mu$m gap in a
thin tin film about 1000 {\AA} thick. A whisker was attached to
electrodes by electrostatic forces at the initial moment, then,
apparently, by the Van der Waals forces. It is not easy to remove
a whisker from the substrate surface. The heat-sinking conditions,
probably, were fairly good because the greater part of the sample
was in contact with the polished substrate surface, therefore
measured CVCs did not exhibit a notable hysteresis in the studied
temperature range, unlike CVCs reported in Refs.
\cite{c13,c14,c15}. CVCs were measured using either the
two-terminal configuration (this was possible because $T_{c}$ of
films was higher than that of whiskers) or the four-terminal
configuration. The substrate supporting the whisker was placed in
a copper waveguide and insulated from environment by a
superconducting lead shield. The curves of the critical current
and resistance versus temperature for the case of a single
phase-slip center at $T_{C}-T<10$ mK had shapes typical of
whiskers \cite{c13}: $I_{c} \sim (1-T/T_{c})^{3/2}$, $R_{0} \sim
(1-T/T_{c})^{-1/4}$ .

\section{EXPERIMENTAL RESULTS}
Current-voltage characteristics of all samples are piece-wise
linear, i.e., they are composed of linear sections connected by
nonlinear sections with larger slopes.
\begin{figure}
\includegraphics[width=1.0\linewidth]{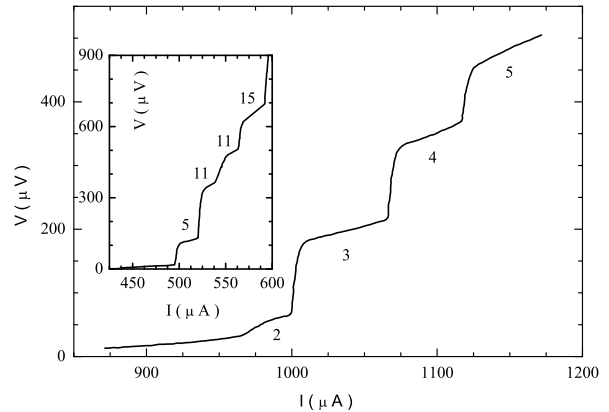}
\caption{\label{fig:1} CVC of the Sn3 whisker ($R_{0} \approx
0.19$ $\Omega$, $T_{c} \approx 3.71$ K, $R_{300}/R_{4.2} \approx
73$) without irradiation by an external microwave field at $T
\approx 3.56$ K. The insert shows the CVC of the Sn2 sample
($R_{0} \approx 0.21$ $\Omega$, $T_{c} \approx 3.72$ K, $d \approx
0.8$ $\mu$m, $R_{300}/R_{4.2} \approx 50$) without irradiation at
$T \approx 3.63$ K.}
\end{figure}
\begin{figure}
\includegraphics[width=1.0\linewidth]{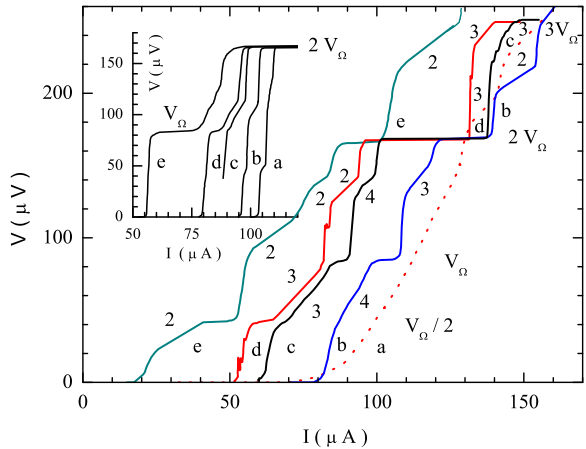}
\caption{\label{fig:2} Set of CVCs of the Sn1 sample
($R_{0}=0.79-0.63$ $\Omega$, $T_{c} \approx 3.69$ K, $d \approx
0.3$ $\mu$m, $R_{300}/R_{4.2} \approx 20$) at different powers of
external microwave irradiation at frequency $\Omega /2 \pi=40.62$
GHz and $T \approx 3.62$ K: (a) 70 dB (dashed line); (b) 36 dB;
(c) 31 dB, (d) 30 dB, (e) 28.6 dB. The insert shows low-current
sections of CVCs of the Sn1 sample at approximately equal
parameters in another cycle of measurements: (a) 32.6 dB; (b) 30.6
dB; (c) 29 dB; (d) 28 dB; (e) 25.2 dB.}
\end{figure}
Figure 1 shows the examples of CVCs of superconducting whiskers
with microwave radiation off. The initial parts of the whisker
CVCs without microwaves are also shown in Figs. 2-4 by dashed
lines. The numbers near the linear sections of the whisker CVCs
indicate the ratios between their resistances and that of a single
phase-slip center, $R_{0}$. The latter parameter was determined as
the largest common divisor of differential resistance values of
all linear CVC sections and compared to an estimate derived from
the size and resistivity of the whisker. The CVC linear sections
are connected by nonlinear sections, which are reproducible and
reversible in the range of studied frequencies. Note that in most
experiments, the initial CVC sections at $I>I_{c}$ (curves (a) in
Figs. 2 and 3) without radiation are nonlinear, and the first
reproducible linear sections correspond to states with several
phase-slip centers (the linear section $3R_{0}$ on curve (a) in
Fig. 2 and $5R_{0}$ on curve (a) of Fig. 3). In earlier
experiments \cite{c13,c14,c15}, the states with one phase-slip
center could be regularly produced. In contrast to those
experiments, where the temperature difference $T_{c}-T$ was less
than 10 mK, we measured CVCs mostly at temperatures 70-160 mK
below $T_{c}$. In this case, states with several phase-slip
centers were stable at notably larger temperature differences
$T_{c}-T$. This can be seen by comparing the CVC shown in Fig. 1
with the CVC in the insert to this graph. Moreover, the linear
sections with the same resistance (such as $3R_{0}$ sections on
curve (a) in Fig. 2 for Sn1 and $2R_{0}$ in curve (d) in Fig. 4
for Sn3) separated by voltage jumps were recorded many times. On
the basis of these observations, we have come to the conclusion
that, in spite of some complications in interpreting CVCs of our
whiskers, they are superconducting channels with phase-slip
centers at appropriate temperatures and transport currents.

When samples are exposed to microwave radiation, their CVCs
contain, in addition to linear sloping sections due to the
presence of certain numbers of phase-slip centers, steps with a
zero slope at voltages
\begin{equation}
U_{m/n}=(m/n)U_{\Omega} \;. \nonumber
\end{equation}
\begin{figure}
\includegraphics[width=1.0\linewidth]{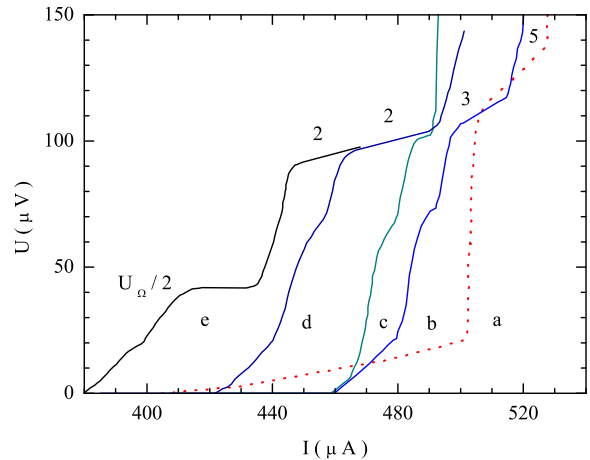}
\caption{\label{fig:3} Set of CVCs of the Sn2 sample
($R_{0}=0.23-0.18$ $\Omega$, $T_{c} \approx 3.72$ K, $d \approx
0.8$ $\mu$m, $R_{300}/R_{4.2}  \approx 50$) at different powers of
microwaves at frequency $\Omega /2 \pi = 40.62$ GHz at $T \approx
3.63$ K: (a) 70 dB (dashed line); (b) 30.2 dB; (c) 30 dB; (d) 28.2
dB; (e) 26.1 dB.}
\end{figure}
\begin{figure}
\includegraphics [width = 1.0\linewidth]{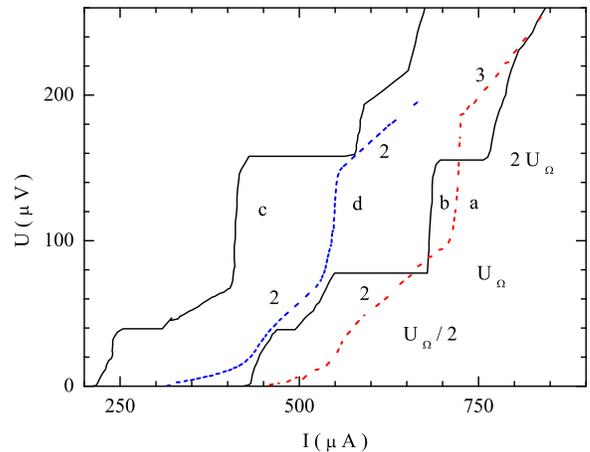}
\caption{ \label{fig:4} CVC of the Sn3 whisker ($R_{0}=0.21$
$\Omega$, $T_{c} \approx 3.71$ K, $R_{300}/R_{4.2} \approx 73$) at
different microwave powers at frequency $\Omega /2 \pi =37.5$ GHz
at $T \approx 3.58$ K (curves a, b, and c) and $T \approx 3.63$ K
(curve d); (a) 70 dB (dashed line); (b) 19.5 dB; (c) 12 dB; (d) 70
dB (dashed line).}
\end{figure}
At low microwave powers, the channel critical current was higher,
i.e., stimulation of superconductivity theoretically described by
Eliashberg \cite{c18} took place. Instead of the emergence of the
zero-slope step first at $U_{\Omega}$ \cite{c13,c14,c15}, we
observed the sequential appearance of steps at $2U_{\Omega}$,
$U_{\Omega}$, $3U_{\Omega}$, and $U_{\Omega}/2$ for a sample Sn1
(Fig. 2), and in Sn2 (Fig. 3) we first observed a step at
$U_{\Omega}/2$ and then at $U_{\Omega}$ (not shown in the graph).
At lower temperatures the unusual shapes of the CVCs at zero
radiation intensity with linear sections of the same slope (curve
(d) in Fig. 4) or the lowest linear sections corresponding to
several phase-slip centers were replaced by more common CVC
shapes. The sequence of microwave-induced steps in whiskers' CVCs
emerging with increasing microwave power also became more like the
usual sequence at lower temperatures, namely, the step at
$U_{\Omega}$ was detected first, then the step at $2U_{\Omega}$,
and at still higher microwave power at $U_{\Omega}/2$ (Fig. 4).
The curve became similar to those given in Refs.
\cite{c13,c14,c15}. As the microwave power increased, the sloping
linear sections due to the phase-slip centers became more
pronounced on CVCs (Figs. 2 and 3).

Steps with zero slope emerge on linear sections of CVCs, which
either exist in the samples not exposed to microwaves or appear in
the samples irradiated by the microwave field. For example, the
step of zero slope on the CVC of the Sn1 whisker at $U_{\Omega}$
(insert to Fig. 2) appears after the emergence of a linear section
on the curve, its growth, and the shift of its lower edge to the
required voltage (curves (a), (b), and (c) in the insert to Fig.
2). As soon as the edge of the linear section achieves
$U_{\Omega}$, a zero-slope step is produced (curve (d), and its
width increases with the microwave power curve (e)). The steps at
$U_{\Omega}/2$ (curves (c), (d), and (e) in Fig. 2) emerge in a
similar manner. The step at $3U_{\Omega}$ (curve (c) in Fig. 2)
appears when the sloping linear section with differential
resistance $3R_{0}$ extends to this region. A zero-slope step can
disappear at a higher microwave power (for example, the steps at
$3U_{\Omega}$, $U_{\Omega}$, and $U_{\Omega}/2$) when the upper
edge of the linear section shifts below the respective voltage,
and a vertical CVC section moves to this region. In this case, the
differential resistance of the linear section can have a jump
(curves (d) and (e) in Fig. 2), namely, the linear section at
about $3U_{\Omega}$ changed its factor from 3 to 2.
\begin{figure}
\includegraphics[width=1.0\linewidth]{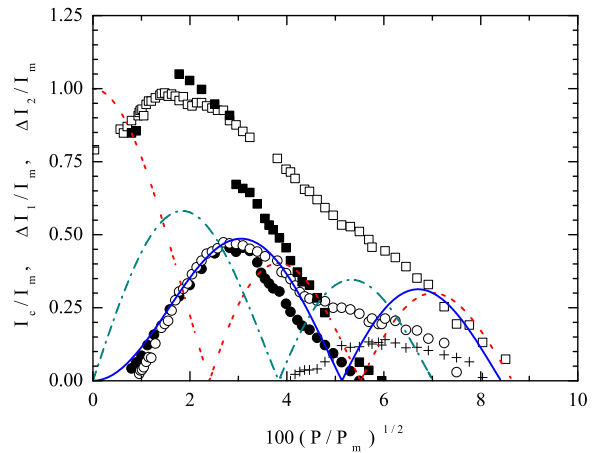}
\caption{\label{fig:5} Normalized critical current $I_{c}/I_{m}$
versus the relative amplitude of external microwave field at
frequency $\Omega /2 \pi=40.62$ GHz at $T \approx 3.62$ K for the
Sn1 whisker in two different cycles of measurements (full squares
are the data of the first cycle and empty squares correspond to
the second cycle), $I_{m} \approx 107$ $\mu$A. Normalized widths
of zero-slope steps on a CVC as functions of relative microwave
field amplitude at voltage $U_{\Omega}$ (crosses plot data of the
second cycle) and at voltage $2U_{\Omega}$ (full circles are the
data of the first cycle and empty circles correspond to the second
cycle). The dashed, dash-dotted, and solid lines show absolute
values of Bessel functions $J_{0}(x)$, $J_{1}(x)$, and $J_{2}(x)$,
respectively; $x=100(P/P_{m})^{1/2}$, $P$ is the power, and
$P_{m}$ is the maximal output of the microwave generator.}
\end{figure}
Thus, a linear section on a CVC of a sample with or without
microwave pumping at $U_{m/n}$ is a necessary condition for
formation of a zero-slope step, i.e., for the existence of the
required number of phase-slip centers in the sample.

By tuning the incident microwave frequency $\Omega$, we could
detect zero-slope steps not observed previously when voltage
$U_{m/n}$ coincided with a linear section of a CVC recorded
without irradiation.

Sloping linear sections in a CVC of a whisker containing a certain
number of phase-slip centers and exposed to microwaves of a fixed
power could decrease their resistance factor with respect to the
resistance of an isolated phase-slip center if the direct
transport current increased (see curve (b) in Fig. 2). The
resistance factor could also remain unchanged (curve (e) in Fig.
2, section 2). An increase in the incident microwave power could
cause, in addition to suppression of both the critical and excess
current at a fixed voltage, a switch-over to a linear section with
a lower differential resistance. On the curves in Fig. 2, the
resistance factor dropped from four to two, and in Fig. 3 from
five to two. The CVCs of the Sn2 whisker (Fig. 3) initially
contained a linear section with resistance $5R_{0}$ at voltages
above $U_{\Omega}$, and under microwave irradiation this parameter
dropped to $3R_{0}$ and then $2R_{0}$. At higher microwave powers
the length of the $2R_{0}$ section increased at a constant
resistance factor. Note that $R_{0}$ could vary under microwave
radiation within 20\% . Thus, microwaves not only produce
horizontal steps on CVCs, but also strongly affect CVCs of tin
whiskers.

We have also measured the widths of microwave induced steps as
functions of the incident power over the interval of their
existence. The experimental dependencies of current-normalized
widths of zero-slope steps at voltages $U_{\Omega}$ and
$2U_{\Omega}$, and of the critical current for a sample Sn1
obtained in different measurement cycles at approximately equal
parameters as functions of the relative microwave amplitude are
given in Fig. 5. The graph also shows as an illustration the
absolute values of Bessel functions of order 0, 1, and 2
($J_{0}(x)$, $J_{1}(x)$, and $J_{2}(x)$) although we believe that
the experimental curves are not directly related to these
functions. Note the main features of the curves in Fig. 5. (1) The
microwave stimulation of superconductivity led to an increase in
the critical current of about 20\% . (2) The zero-slope step at
$2U_{\Omega}$ emerged at a lower microwave power and had the
maximum width of about $0.5I_{c}$. (3) The step at $U_{\Omega}$
observed in the second cycle of measurements (it was too small in
the first cycle and its width is not shown in Fig. 5) appeared at
a higher microwave power, and in its presence the width of the
$2U_{\Omega}$ step and the critical current as functions of the
microwave field amplitude changed considerably. In this case the
critical current and width of the $2U_{\Omega}$ step vanished at a
notably higher microwave field amplitude than in the first cycle.
(4) Induced $U_{\Omega}$ and $2U_{\Omega}$ steps appeared at a
finite microwave power, i.e., there is a certain threshold
microwave power needed for formation of these steps. This
threshold is related to the extension of the linear CVC sections
to voltages $U_{\Omega}$ and $2U_{\Omega}$. (5) There is only one
interval of the microwave field amplitude on which the critical
current and CVC steps exist. No oscillations have been detected on
the curves of critical current and step width.

In studying step widths as functions of the microwave power, we
recorded (in several cases) nonmonotonic curves with relatively
narrow down-peaks against the background of wide bell-shaped
curves.

\section{DISCUSSION OF RESULTS}
The current-voltage characteristic of a uniform superconducting
channel, which is our model for a whisker, depends on its length.
In the case of a short whisker section through which current is
fed, $l \approx l_{E}$, the presence of one phase-slip center
allows the sample to conduct a current higher than the critical
value. If $l \gg l_{E}$, the exponentially decaying parameters of
phase-slip centers have little effect on the channel properties,
therefore it should contain several phase-slip centers, whose
number is determined by the channel length. In our samples, the
condition $l \gg l_{E}$ was satisfied ($l \approx (10-20)l_{E}$),
therefore we assume that several phase-slip centers were necessary
to conduct a current slightly higher than the critical value. In
the process of generation of the required number of centers, the
instantaneous number of centers can be unstable and variable in
both time and space.

The CVCs of our samples have piecewise linear shapes with sections
characterized by differential resistance $R=nR_{0}$, where $n$ is
an integer. These sections correspond to definite numbers of
phase-slip centers, which can be derived from the sample sizes. In
addition, there are the nonlinear sections on which the number of
centers is probably unstable and varies with time. A dedicated
investigation is needed to verify this hypothesis. The linear
sections of CVCs of the superconducting channel in the simplified
model \cite{c7} are described by the formula
\begin{equation}
U=nR_{0}(I-I_{0}) \;. \nonumber
\end{equation}
The excess current $I_{0}$ is usually related to the average
superconducting component of the total current. This formula is
not universal for all linear sections, because $I_{0} \neq const$
for all groups of linear sections \cite{c9}. The CVCs of our
samples contain neighboring linear sections with equal $n$ but
different $I_{0}$.

Microwave irradiation of our samples has a dual effect on their
CVCs. The first effect is the generation of constant voltage
steps, which was the main subject of the reported study. The
second effect is the change in the number of phase-slip centers
under microwave radiation and stabilization of CVC sections with
definite numbers of these centers. This shows up in the extension
of linear sections and transformation of some nonlinear CVC
portion to linear.

The existence of constant-voltage steps under microwave radiation
indicates that there are currents of microwave frequencies with
spectral components \[ \omega=2enU_{\omega}/\hbar \;,
n=1,2,3,...\] in the regions of phase-slip centers. When the
external frequency equals that of one of these harmonics, several
centers are synchronized, which shows up in the form of
constant-voltage steps at
\[ U_{m}= mU_{\omega} \;, \] where $m$ is the
number of phase-slip centers, \[U_{\omega} = \hbar \Omega /2en \;,
\] and $\Omega$ is the external radiation frequency.
\begin{figure}
\includegraphics[width=1.0\linewidth]{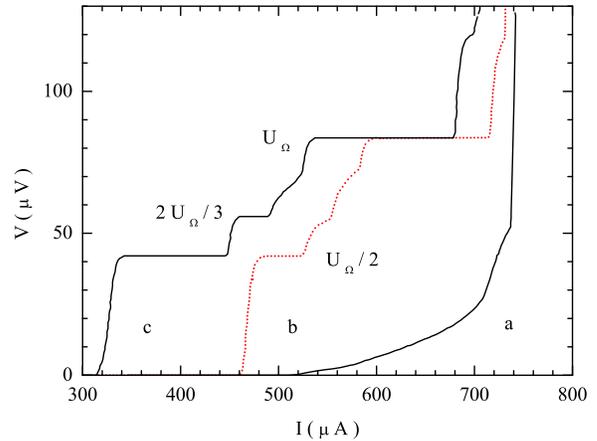}
\caption{ \label{fig:6} Low-current sections of CVCs of the Sn2
sample at different microwave field powers at frequency $\Omega /2
\pi=40.62$ GHz in the second cycle of measurements at $T \approx
3.60$ K: (a) 70 dB; (b) 24.49 dB; (c) 22 dB.}
\end{figure}
As a result, steps can occur at \[ U_{m/n}=(m/n) \hbar \Omega /2e
\] if this voltage coincides with an inherent or microwave-induced
linear section of CVC with a definite (integral) number of
phase-slip centers.

Unfortunately, it is difficult to determine $m$ and $n$ with
certainty using CVCs. Linear sections from which the number of
centers could be exactly determined could be seen near
constant-voltage steps only at certain values of parameters. We
believe that the step at $U_{\Omega}/2$ in Fig. 2 is due to the
synchronization of two phase-slip centers by the fourth harmonic
of proper oscillations, i.e., $U_{\Omega}/2 \to 2U_{\Omega}/4$,
similarly $U_{\Omega} \to 4U_{\Omega}/4$, $2U_{\Omega} \to
4U_{\Omega}/2$, 3$U_{\Omega} \to 6U_{\Omega}/2$, the step at
$4U_{\Omega} \to 8U_{\Omega}/2$ is not shown; in Fig. 4
$U_{\Omega}/2 \to 2U_{\Omega}/4$, $U_{\Omega} \to 2U_{\Omega}/2$,
and $2U_{\Omega} \to 4U_{\Omega}/2$; in Fig. 3, $U_{\Omega}/2 \to
2U_{\Omega}/4$; in Fig. 6, $U_{\Omega}/2 \to 3U_{\Omega}/6$,
$2U_{\Omega}/3 \to 4U_{\Omega}/6$, and $U_{\Omega} \to 6U_{\Omega}
/6$. At other values of parameters this sample demonstrated steps
at $5U_{\Omega}/6 \to 5U_{\Omega}/6$ and $U_{\Omega}/3 \to
2U_{\Omega}/6$ (not shown in the graphs of this paper).

Thus, at certain positions of these voltages in the whisker CVCs,
microwave field synchronizes oscillations of the order parameter
in all phase-slip centers present in a sample, which results in
constant voltage drops across isolated centers and across the
entire sample. States with synchronized phase-slip centers under
microwave radiation emerge predominantly at corresponding
locations in the CVCs. Other CVC sections may correspond to states
in which some phase-slip centers are synchronized by external
field and the rest are not. This conjecture allows us to interpret
the drop in the differential resistance of linear sections (and
the behavior of the differential resistance in general) when the
current increases under microwave radiation. The existence of
neighboring sloping steps with equal resistance but different
excess current can also be interpreted in these terms. A similar
effect without microwave radiation can be attributed to a
different but, in a sense, similar phenomenon. So-called Fiske
steps \cite{c19} were detected in experiments with tunneling
Josephson junctions when the frequency generated by the junction
was locked to the resonant frequency of the structural cavity in
the experimental device. In this case, constant-voltage steps
determined by the Josephson formula with the resonant cavity
frequency could be seen on CVCs. The gap in the tin film on which
the whisker was mounted could act as a structural resonator. The
length of this gap was about 5 mm, and, given the silicon
substrate dielectric constant ($\varepsilon \approx 12$), we have
a resonant frequency in the studied microwave band. In this case,
a section with a constant voltage due to synchronization of a
group of phase-slip centers (Fiske step) can occur. The centers
whose oscillations are not locked to the resonant frequency should
demonstrate a linear behavior. As a result, the CVC of the sample
should have a linear section with the resistance corresponding to
the number of unlocked centers, which is smaller than the total
number. The question why horizontal steps have not been observed
remains unanswered. Doubts in this interpretation could be
eliminated by directly measuring microwaves generated in the
sample.

Figure 5 shows the widths of constant-voltage steps as functions
of the microwave field amplitude in relative units. The maximal
width of these steps allows us to estimate the microwave power
generated by the whisker:
\begin{equation}
P \approx (\Delta I)^{2}mR_{0} \;, \nonumber
\end{equation}
where $\Delta I$ is the step width in terms of current. Hence,
$P\approx 10^{-8}$ W.

The microwave generation in the phase-slip centers can be
interpreted in terms of the order parameter versus time, which
vanishes at some moment and then increases to some value. At the
moment when the order parameter is zero, the difference between
the phases on different sides of the phase-slip center drops by
$2\pi$. It would be interesting to estimate the times of these
processes and compare their reciprocal values with the frequencies
of the order parameter oscillations and external radiation. The
most important parameter is the time $\tau_{\Delta}$ in which the
order parameter recovers. When $\tau_{\Delta}$ is much longer than
the order parameter oscillation period determined by the Josephson
formula, both the mean and instantaneous absolute values of the
order parameter within the center are much smaller than the
equilibrium value in other regions of the superconducting channel.
If $\tau_{\Delta}$ is comparable to or smaller than the period of
the order parameter oscillations, the instantaneous value of the
gap in the phase-slip center can be large and comparable to the
gap in the surrounding regions. The spectra of normal excitations
in phase-slip centers should be notably different in these two
cases, which can lead to differences in some electrical properties
of phase-slip centers. Since the energy relaxation time of current
carriers in tin is $3 \times 10^{-10}$ s, the first case is
realized in the microwave frequency band.

The behavior of the step width is determined by two factors. The
first is the width of the step against the background of an
infinite linear CVC section with a definite number of phase-slip
centers as a function of the microwave field amplitude. The second
is the limitation of the constant-voltage step by the length of
the CVC linear sloping section, whose positions, as follows from
experimental data, are also functions of the microwave power. A
change in the number of phase-slip centers breaks the initial
synchronization condition, and the system can switch to either a
totally un-synchronized state, or a partially synchronized state,
or fully synchronized state at a different harmonic and with a
different number of phase-slip centers (for example, the
zero-slope step at $2U_{\Omega}$ in Fig. 2 can be due to
synchronization of four centers by the second harmonic or six
centers by the third harmonic). Given these two effects, we could
not determine the constant-voltage step width as a function of the
microwave amplitude unambiguously and compare it to the
theoretical model. The existence of the microwave power threshold
at which induced steps appear and the absence of oscillations in
both the zero-slope step width at $U_{m/n}$ and critical current
as functions of the microwave field amplitude are due to a
definite number of phase-slip centers required at these voltages.

\section{CONCLUSION}
In the reported work, we have studied the effect of microwave
radiation on current-voltage characteristics of whiskers with
submicron diameters. Such whiskers can serve as microwave
oscillators at frequencies of up to 40 GHz with an output of about
$10^{-8}$ W. The spectrum of generated waves contains many
harmonics, and the generation occurs on CVC sections with stable
numbers of phase-slip centers. Features of CVCs of our samples
under microwave radiation are determined by changes in the number
of phase-slip centers and the synchronization degree of generation
in these centers.

The work was supported by the \textit{Superconductivity}
subprogram of the \textit{Physics of Condensed State} program
sponsored by the Russian government (Project No. 95021), and by
the \textit{Physics of Solid-State Nanostructures} program
(Project No. 1-084/4).


\begin{thebibliography}{99}

\bibitem{c1} D. W. Palmer and J. E. Mercereau, Appl. Phys. Lett. \textbf{25}, 467
(1974).

\bibitem{c2} M. Octavio and W. J. Skocpol, J. Appl. Phys. \textbf{50}, 3505
(1979).

\bibitem{c3} L. E. Amatuni, V. N. Gubankov, A. V. Zaitsev,
and G. A. Ovsyannikov, Zh. Eksp. Teor. Fiz. \textbf{83}, 1851
(1982) [Sov. Phys. JETP \textbf{56}, 1070 (1982)].

\bibitem{c4} L. E. Amatuni, V. N. Gubankov, and G. A. Ovsyannikov, Fiz. Nizkikh Temp. \textbf{9},
939 (1983) [Sov. J. Low Temp. Phys. \textbf{9}, 484 (1983)].

\bibitem{c5} B. I. Ivlev
and N. B. Kopnin, Usp. Fiz. Nauk \textbf{142}, 435 (1984) [Sov.
Phys. Usp. \textbf{27}, 206 (1984)].

\bibitem{c6} R. Tidecks, \textit{Current-Induced Nonequilibrium
Phenomena in Quasi-One-Dimensional Superconductors,} in
\textit{Springer Tracts in Modern Physics}, Vol. \textbf{121},
Springer (1990).

\bibitem{c7} W. J. Skocpol, M. R. Beasley, and M. Tinkham,
J. Low Temp. Phys. \textbf{16}, 145 (1974).

\bibitem{c8} S. M. Gol'berg, N. B. Kopnin, and M. I. Tribel'skii, Zh. Eksp.
Teor. Fiz. \textbf{94}, 289 (1988) [Sov. Phys. JETP \textbf{67},
812 (1988)].

\bibitem{c9} J. Meyer and G. Minnigerode, Phys. Lett. A \textbf{38}, 529 (1972).

\bibitem{c10} J. D. Meyer and R. Tidecks, Solid State Commun. \textbf{24}, 639 (1977).

\bibitem{c11} M. Tinkham, J. Low Temp. Phys. \textbf{35}, 147 (1979).

\bibitem{c12} X. Yang and R. Tidecks, Z. Phys. B \textbf{83}, 113 (1991).

\bibitem{c13} R. Tidecks and G. von Minnigerode, Phys. Status Solidi A \textbf{52}, 421 (1979).

\bibitem{c14} R. Tidecks and G. Slama, Z. Phys. B \textbf{37}, 103 (1980).

\bibitem{c15} B. Damaschke and R. Tidecks, Z. Phys. B \textbf{77}, 17 (1989).

\bibitem{c16} B. I. Ivlev and N. B. Kopnin, Solid State Commun. \textbf{41}, 107
(1982).

\bibitem{c17} G. E. Churilov, V. M. Dmitriev, and V. N. Svetlov, Fiz. Nizk. Temp.
\textbf{9}, 495 (1983) [Sov. J. Low Temp. Phys. \textbf{9} 250
(1983)].

\bibitem{c18} G. M. Eliashberg, JETP Lett. \textbf{11}, 114 (1970).

\bibitem{c19} M. D. Fiske, Rev. Mod. Phys. \textbf{36}, 221 (1964).

\end{thebibliography}
\end{document}